\documentclass[
twocolumn,
superscriptaddress,
amsmath,amssymb,
aps,
prl,
floatfix,
nofootinbib,
longbibliography]{revtex4-1}
\usepackage{color}
\usepackage{xcolor}
\usepackage{graphicx}
\usepackage{dcolumn}
\usepackage{bm}
\usepackage{ulem} 
\usepackage{cancel}
\usepackage{soul}
\newcommand{\ket}[1]{\lvert #1\rangle}
\newcommand{\bra}[1]{\langle#1 \rvert}

\newcommand{\expect}[1]{\langle #1\rangle}

\begin{document}
	
	\title{Blocking-state influence on shot noise and conductance in quantum dots}
	
	\author{M.-C.~Harabula}
	\email{cezar.harabula@unibas.ch}
	\affiliation{Department of Physics, University of Basel, Klingelbergstrasse 82, CH-4056 Basel, Switzerland}
	
	\author{V.~Ranjan}
	\affiliation{Department of Physics, University of Basel, Klingelbergstrasse 82, CH-4056 Basel, Switzerland}
	\affiliation{
		Quantronics Group, SPEC, CEA, CNRS, Universit\'e Paris-Saclay, CEA Saclay, F-91191 Gif-sur-Yvette, France
	}

	\author{R.~Haller}
	\author{G.~F\"ul\"op}	
	\author{C.~Sch\"onenberger}
	\affiliation{Department of Physics, University of Basel, Klingelbergstrasse 82, CH-4056 Basel, Switzerland}

	\date{\today}
	
	\begin{abstract}
	Quantum dots (QDs) investigated through electron transport measurements often exhibit varying, state-dependent tunnel couplings to the leads. Under specific conditions, weakly coupled states can result in a strong suppression of the electrical current and they are correspondingly called blocking states. Using the combination of conductance and shot noise measurements, we investigate blocking states in carbon nanotube (CNT) QDs. 
	We report negative differential conductance and super-Poissonian noise. The enhanced noise is the signature of electron bunching, which originates from random switches between the strongly and weakly conducting states of the QD. Negative differential conductance appears here when the blocking state is an excited state. In this case, at the threshold voltage where the blocking state becomes populated, the current is reduced.
	Using a master equation approach, we provide numerical simulations reproducing both the conductance and the shot noise pattern observed in our measurements.

	\end{abstract}

	\maketitle
		
	\section{Introduction}
	Beyond time-averaged current, the measurement of current fluctuations gives insight into the interaction and correlations of charge carriers in mesoscopic transport \cite{Blanter2000}. 
	Current fluctuation arising from the quantized nature of charge carriers is called shot noise. Non-interacting, independent particles exhibit Poissonian shot noise with spectral density $S_{\rm Poisson}=|2 e\expect{I}|$, also called Schottky noise, where $e$ is the charge of the carriers and  $\expect{I}$ is the time-averaged current.
	In general, interactions result in correlations and the suppression or enhancement of shot noise. The Fano factor, defined by $F=S/|2 e\expect{I}|$ measures this modification and the granularity of the current: sub-Poissonian noise ($F<1$) is characteristic for anti-bunched charge carriers, while super-Poissonian noise ($F>1$) for bunched transport.
	
	Electrical transport and noise phenomena in quantum dots (QDs) have been studied in experimental \cite{Okazaki2013,Gustavsson2006,Zarchin2007,Onac2006,Ferrier2015} and theoretical works \cite{Sukhorukov2001,Hanke93,Hershfield93,Chen92}. Theory has shown that in single-level QDs the Pauli exclusion principle and the repulsive Coulomb interaction result in anti-bunching \cite{Sukhorukov2001,Hanke93,Hershfield93,Chen92}. However, occupation dynamics in multi-level QDs can give rise to bunching, and correspondingly, super-Poissonian noise \cite {Thielmann2005, Belzig2005, Carmi2012}.
	The electron transport in an interacting two-level system is a telegraphic process if the tunnel couplings of one level are much stronger than of the other \cite{Carmi2012}. This system supports a high current through the strongly coupled level, which is strongly reduced for random intervals when the weakly coupled level is filled.  The electrons transferred in the highly conducting state form bunches and result in enhanced noise. Recently, noise measurements have been applied to probe the correlations induced by the many-body Kondo effect \cite{Ferrier2015}. It has been demonstrated that the increased effective charge results in enhanced shot noise.
	
	In general, such a state in which the QD can be trapped, thus blocking the current, is referred to as a blocking state. We investigate blocking states through conductance and noise spectroscopy in CNT QDs.
	We encounter strongly enhanced noise (up to $F\approx8$) in different transport regimes, inside and outside Coulomb diamonds. Although the details of the underlying microsopic processes are different, both cases present telegraphic transport, induced by asymmetric tunnel couplings.
	After providing a qualitative explanation of the observations, we employ the master equation framework developed in Ref.~\cite{Kaasbjerg2015} to reproduce the conductance and noise pattern seen in our experiments.
	
	\section{Experimental results}
	We realize two devices, each one consisting of a quantum dot in a semiconducting carbon nanotube, in order to measure their current noise at microwave frequencies. Device A (B) utilizes an on-chip lumped $LC$ circuit \cite{Harabula2017} (stub tuner~\cite{Hellmann2012,Hasler2015}) for transforming the  high QD resistance $R$ into a value close to the characteristic impedance of the microwave elements, $Z_0=50~\Omega$. The simplified measurement setup and a scanning electron microscope (SEM) image of a typical device are shown in Fig.~1(a).  The fabrication starts with depositing a 100-nm layer of Nb on an undoped, oxidized silicon substrate. The matching circuit of device A (B) is defined by patterning the Nb layer with $e$-beam (UV) lithography and reactive-ion etching with Ar-Cl$_2$ plasma \cite{Harabula2017,Hasler2015}. To avoid the detrimental effects of CNT growth on the microwave properties of the matching circuit, the CNTs are grown on a donor substrate and transferred onto the device substrate by stamping \cite{Viennot2014}. Further, the selected CNT for device A (B) is contacted with ohmic Ti/Au (Pd/Al\footnote{The proximity gap induced by the superconducting Al lead is much smaller than the bias voltage range used in the experiments and does not play any role in the noise features.}) electrodes. In the same  step, a side gate is evaporated for tuning the electrochemical potential of the QD. 
	
	\begin{figure}[b!]
		\includegraphics[width=\columnwidth]{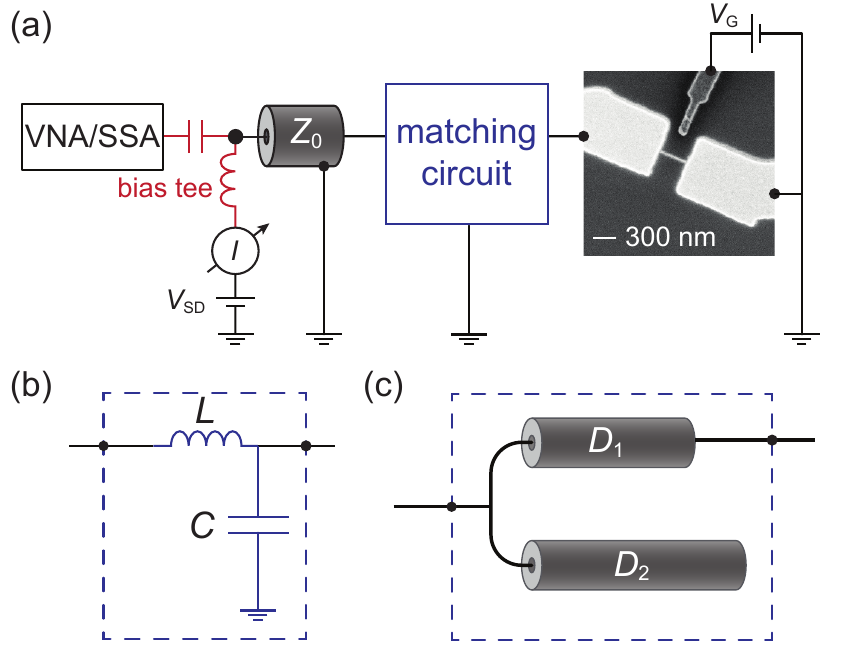}
		\caption{Measurement setup. (a) A SEM image of CNT device connected to a GHz impedance-matching circuit for efficient collection of noise signals. A bias tee enables simultaneous measurement of dc and rf properties. The source electrode is coupled to the on-chip matching circuit, while the drain is connected to the ground plane. Schematics of (b) a lumped $LC$ impedance transformer, used in device A and (c) a stub tuner based on coplanar transmission lines, used in device B. Both matching circuits are fabricated from Nb films and have a resonance frequency of $\approx 3$~GHz.}
		\label{fig:setup}
	\end{figure}
	
	Both matching circuits behave as band pass filters around the resonance frequency $f_0\approx 3$~GHz. At full matching, the two circuits provide the same figure of merit in terms of signal-to-noise ratio for power collected in the entire bandwidth, $\Delta f = \rm FWHM$ ~\cite{Hasler2015}.
	However, $LC$ circuits offer a significantly larger bandwidth 
	($\Delta f_{LC}/f_0 \sim 2 \sqrt{Z_0/R}$) compared to stub tuners ($\Delta f_\mathrm{stub}/f_0=4 \pi ^{-1} Z_0/R$), enabling  much faster acquisition of signals.
	
	The measurements are performed in a dilution refrigerator at an electronic temperature of $T_e \approx 50$~mK. First, we characterize the matching circuit by fitting the rf reflectance curve \cite{Harabula2017,Hasler2015}, measured with a vector network analyzer (VNA). Second, the QD differential conductance $G=R^{-1}$ is calculated from rf reflectance \cite{Hasler2015,Harabula2017}. Third, the QD current noise $S_I$ is extracted from the transmitted noise power, measured with a signal and spectrum analyzer (SSA) \cite{Hasler2015,Harabula2017}. 
	In the $LC$ circuit case, we set a acquisition bandwidth of $\Delta f_{{\rm a,} LC}=50$~MHz \cite{Harabula2017}; in the stub tuner case, we choose  $\Delta f_{\rm a, stub}=5$~MHz. Simultaneously, the time-averaged current $\langle I \rangle$ is recorded with dc readout.

	\begin{figure*}
		\includegraphics[width=\textwidth]{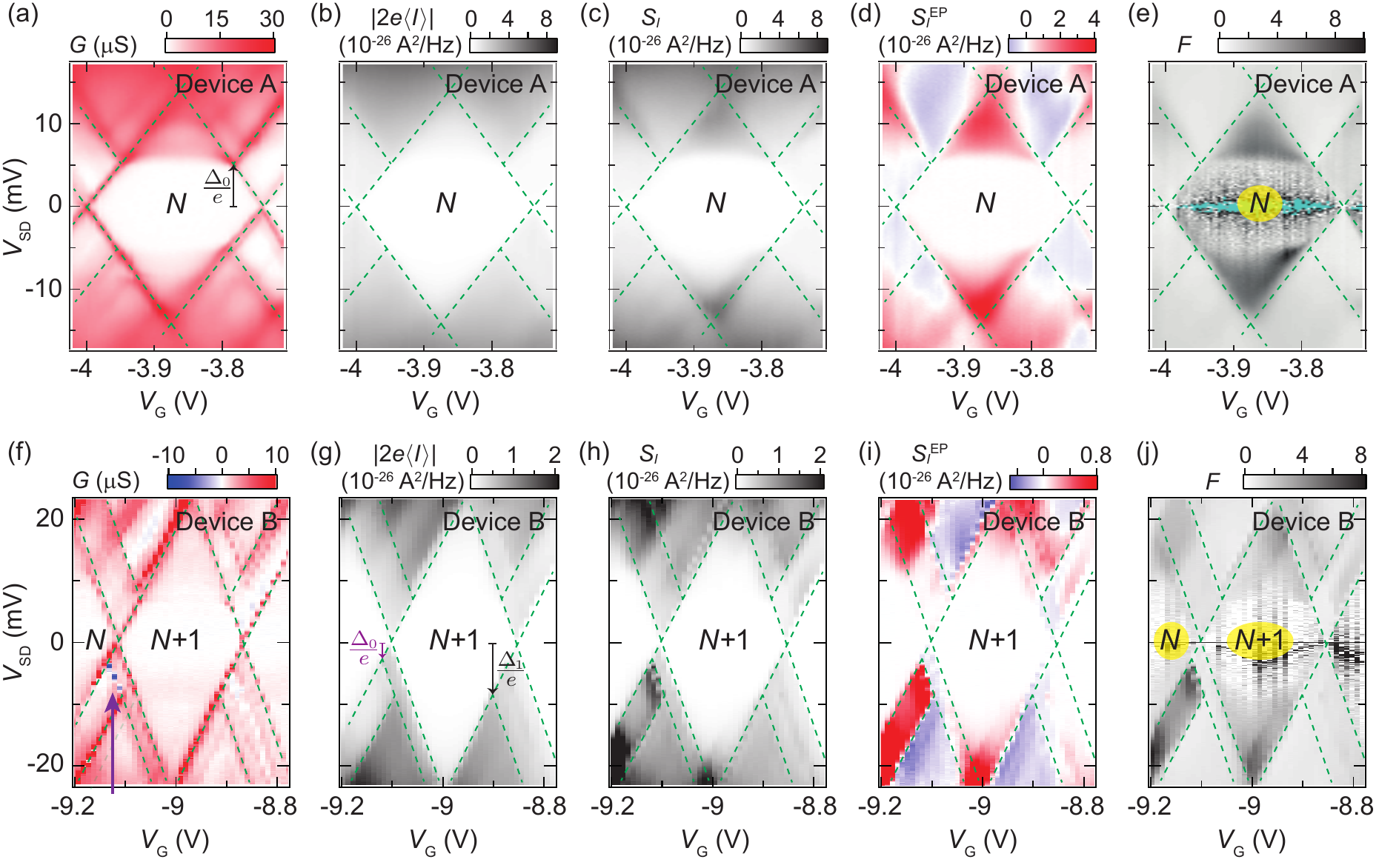}
		\caption{(Color online) Conductance and noise measurements of device A (a-e) and device B (f-j). (a),(f) Differential conductance deduced from  RF reflectance, as a function of bias voltage $V_{\rm SD}$ and side gate voltage $V_{\rm G}$. Dashed green lines serve as guides for the eye; the same guides are overlaid in the remaining panels. The purple arrow in (f) points to a segment of negative differential conductance. (b),(g) The Schottky noise, $|2e \langle I\rangle|$, provide a comparison reference for current noise. (c),(h) Measured current noise spectral density, $S_I$. (d),(i) The excess noise, $S_I^\mathrm{EP} = S_I-|2e\langle I \rangle|$. (e),(j) The Fano factor, obtained by dividing the current noise spectral density $S_I$ by the simultaneously measured Schottky value $|2e\langle I\rangle|$. Cyan areas inside the diamonds represent divergent domains, where uncertain noise values are divided by small currents. }
		\label{fig:data}
	\end{figure*}
	
	The data of device A are plotted in Figs.~2(a)-(e). Specifically, Fig.~2(a) shows the differential conductance, as a function of bias voltage, $V_{\rm SD}$, and side gate voltage, $V_{\rm G}$. Figs.~2(b) depicts the noise spectral density reference, $S_{\rm Poisson} = |2e\expect I |$. This reference is calculated by scaling the measured current $\expect I$ with $2e$. Fig.~2(c) presents the measured noise spectral density, $S_I$, of the QD. Further, the enhanced noise is illustrated in two ways: the QD excess noise, $S_I^{\rm EP} = S_I- |2e\expect I |$, appears in Fig.~2(d) and the Fano factor, $F = S_I / |2e\expect I |$, is mapped in Fig.~2(e). Analogously, the data of device B are plotted in Figs.~2(f)-(j).
	
	Guides for the eyes highlight the main lines of the differential conductance maps. Both measurements show Coulomb diamonds, from the height of which we extract a charging energy of $U_c \approx 15$~meV (device A) and $U_c \approx 20$~meV (device B). The diamonds are labeled with the electron occupation number (e.g. $N$). Current through QDs usually consists of sequences in which one electron tunnels from a lead into the dot, increasing $N$ by one, then tunnels out to the other lead, decreasing the charge of the QD---that is sequential tunneling. Each such electron hopping is called a first-order tunneling event. Inside the diamonds, as first-order tunneling is prohibited, the system is in Coulomb blockade.
	
	Outside the diamonds, the blockade is lifted. Also here, the conductance plots exhibit high-$G$ lines starting at a finite bias: about 5~mV (device A) and approximately 2~mV, 8~mV (device B). These lines originate from excited states at a fixed electronic occupation and their threshold bias corresponds to the excitation energy, provided by the bias voltage:
	$|eV_{\rm SD}|= \Delta_{0,1}$.  We designate excited states with a star superscript, e.g. $\ket {N^*}$. Therefore, $\Delta_0 \approx 5$~meV for state $\ket {N^*}$ in device A and $\Delta_0 \approx 2$~meV, $\Delta_1 \approx 8$~meV for states $\ket {N^*}$, $\ket {(N+1)^*}$ in device B.

	We now focus on device A. Inside Coulomb diamond $N$, in the absence of first-order tunneling, a low current still flows due to second-order tunneling (i.e. cotunneling). Cotunneling means that one charge passes coherently, in one event, through both tunnel barriers of the QD. In this process, the charge state of the QD is preserved. Elastic cotunneling does not change the final state of the QD and is possible at any value of the bias voltage. In contrast, inelastic cotunneling (IEC) alters the QD state, e.g. $N \rightarrow N^*$, the needed energy being provided by the bias voltage, $e|V_{\rm SD}| \geq \Delta_0$. Here, because we observe a conductance increase at the excitation bias, $e|V_{\rm SD}| = \Delta_0$ [Fig.~2(a)], the crucial role is played by IEC. The IEC regime inside the Coulomb diamond appears also in the noise-related maps [Fig.~2(c)-(e)], with a strong Fano factor enhancement of up to $F\approx 8$.
	Outside the Coulomb diamond, below the excited state lines one observes mostly $0.5 < F <1$, as it is expected for transport in a double tunnel barrier \cite{Blanter2000, Hasler2015}.

	In the case of device B, the same Coulomb diamond inner structure can be observed as in device A, but the increase of conductance due to IEC is weaker. Another difference is the very pronounced line of negative differential conductance (NDC) starting from the Coulomb diamond $N$ and here depicted in blue. While typically the current rises when the absolute bias voltage increases and a transition enters the bias window, as in device A, here the current is suppressed. However, the NDC ridge is confined to a segment, between the Coulomb diamond edge and a parallel line of high conductance. 
	Regarding the noise produced in device B, the striking feature is the enhanced $F\approx 6$ just outside Coulomb diamond $N$, at negative bias. By comparing the the conductance and Fano factor maps, one can see that this region of enhanced noise is a band bordered by the NDC ridge and parallel to the Coulomb diamond edge.

	To summarize, our two key findings are: (i) in device A, strongly super-Poissonian noise inside the Coulomb blockade, above a threshold voltage determined by an excited state and (ii) in device B, strongly super-Poissonian noise outside the Coulomb blockade, involving excited states. The origin of these two findings will be detailed in the following. 

	\section{Model}

	\begin{figure*}
	\includegraphics[width=\textwidth]{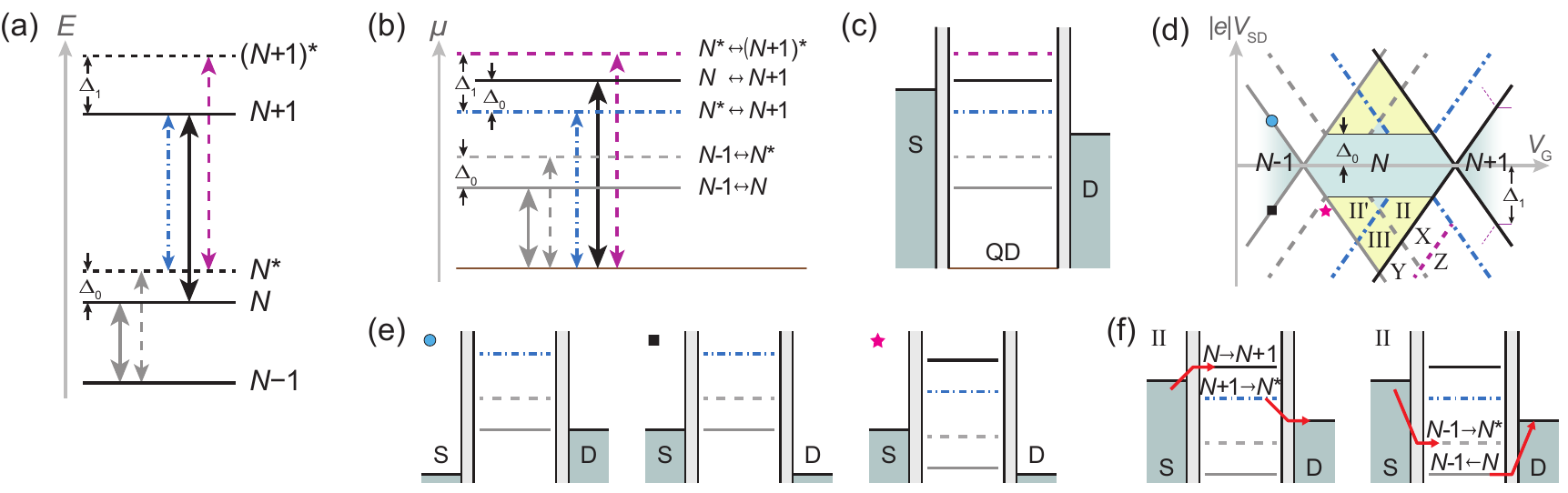}
	\caption{(Color online) (a) Energy levels in the considered model, with transitions marked as dashed arrows (b) Electrochemical-potential levels, given by the arrows in (a) and a common offset. (c) QD transitions represented as electrochemical-potential levels, whose positions are determined in (b). (d) Schematic stability diagram with relevant transitions. Positive- (negative-) slope lines depict resonances of the source (drain) Fermi level with quantum state transitions depicted in (c). (e) Chemical-potential diagrams for the lead-resonant lines [circle and square points in (d)] and for a domain inside the cone of the full gray lines and the cone of the dashed gray lines [star points in (d)]. (f) Two paths for inelastic cotunneling [for region II in (d)].}
	\label{fig:model}
	\end{figure*}

	In this section, we detail a simple model for our QDs, similar to the model used in Ref.~\cite{Kaasbjerg2015}, and show how it is projected onto electrochemical-potential diagrams and further onto charge stability diagrams. Lastly, we place transport mechanisms on the stability diagram.
	
	The QD is described by a spinless model, with states labeled by the electronic occupation number:
	\begin{align}
		\ket{N-1}, ~\ket{N}, ~\ket{N^*}, ~\ket{N+1},~\ket{(N+1)^*}.
	\end{align}
	The star superscript denotes an excited state, e.g. $\ket N$ is a ground state, $\ket{N^*}$ is an excited state, and both states have $N$ electrons.
	The corresponding QD energies are:
	\begin{align}
		E_{N-1} &=0,&&\\ 
		E_N &=\epsilon_0, 	&E_{N^*} &= \epsilon_0 +\Delta_0,& \\ 
		E_{N+1} &=2\epsilon_0+U_{\rm c}, &E_{(N+1)^*}&=  2\epsilon_0 + U_{\rm c}+\Delta_1,& 
	\end{align}
	where $\epsilon_0$ is the kinetic and confinement energy of an additional charge and $U_{\rm c}$ is the charging energy for double occupancy. The excitation energies $\Delta_0$ and $\Delta_1$, different from each other by virtue of many-body effects, are free parameters in our model.
	
	Fig.~\ref{fig:model}(a) illustrates the energy levels (horizontal lines) and possible  transitions between them. Each sketched transition (a dashed vertical arrow) between two states $\ket N$ and $\ket M$ that differ by one electron represents the addition energy $\mu_{N\leftrightarrow M} =|E_N - E_M|$, also known as chemical-potential level. We order these transitions in Fig.~\ref{fig:model}(b), from a common potential offset, which is proportional to $-|e|V_{\rm G}$. The result of their ordering is Fig.~\ref{fig:model}(c), showing the chemical-potential diagram of the QD.
	The transitions pictured in a chemical-potential diagram are involved in first-order tunneling (i.e. electron tunneling through a barrier) and second-order cotunneling (i.e. electron cotunneling, mediated by virtual states). Furthermore, the internal excitation or relaxation are transitions between same-charge states, that do not involve tunneling.

	Depicting transitions in a charge stability diagram is the aim of Fig.~\ref{fig:model}(d). Here, we remind the significance of various lines in the $(V_{\rm G}, V_{\rm SD})$ maps. Remark first the $N$-labeled Coulomb diamond. The underlying lines of this diamond's edges correspond to transitions to and from state $\ket N$: the left (right) edges represent transitions between $\ket N$ and $\ket{N-1}$ ($\ket{N+1}$). All sketched lines mark the alignment (resonance) of a chemical potential [Fig.~\ref{fig:model}(c)] with the Fermi level of a lead: positive-slope lines indicate resonance with source, negative-slope lines indicate resonance with drain, as further exemplified in Fig.~\ref{fig:model}(e). Line graphic styles in Fig.~\ref{fig:model}(d) are replicated from corresponding transitions in Fig.~\ref{fig:model}(c).

	Before placing significant transport mechanisms on the stability diagram, we explain the dashed lines and define several distinct regions.
	Dashed lines correspond to transitions involving the excited state $\ket{N^*}$ and we conveniently call them excited lines. The most typical excited lines are the lead-resonant $N\pm 1 \leftrightarrow N^*$ transitions. Outside the Coulomb diamond, they start at voltage biases $|eV_{\rm SD}| = \Delta_0$. Inside the diamond, together with the horizontal line $|e|V_{\rm SD} = -\Delta_0$, the excited lines form the regions II, II', III \cite{Wegewijs2001}.

	A particular dashed line in Fig.~\ref{fig:model}(d) is the purple one below diamond $N$. It corresponds to the source-resonant $N^* \leftrightarrow (N+1)^*$ transition. This excited line, together with its neighboring gray and blue excited lines, define below diamond $N$ the regions X, Y, Z.  One can establish that the right corner of X is situated at a bias $|e|V_{\rm SD}=-\Delta_{1}$. Indeed, in the charge stability diagram, two lines intersect at a bias given by their level difference in Figs.~\ref{fig:model}(b) or (c).

	The first- and second-order transport processes, and eventually their interplay, are mapped in the following.
	
	For sequential tunneling to occur, a QD transition level is needed in the bias window of the QD, that is between the Fermi levels of the two leads. For instance, Fig.~\ref{fig:model}(e) shows the full gray line in the bias window---this is equivalent, on a charge stability diagram, to the point ($V_{\rm G},V_{\rm SD}$) being in the cone of the two full gray lines. 
	Regions II and III, in the cone of the dashed blue lines, exhibit sequential tunneling through the transition $N^* \leftrightarrow N\!+\!1$. The sequence undergone here by the state of the QD is $N^* \rightarrow N\!+\!1 \rightarrow N^* \rightarrow N\!+\!1 \rightarrow ...$ Region X, entirely laid under the cone of the lead-resonant $N\leftrightarrow N\!+\!1$ full black lines, is another example of domain with possible sequential tunneling.
	
	The second-order process important in noise is the IEC. Figs.~\ref{fig:model}(f) illustrate two possible IEC paths: either (i) one electron tunnels from S into the QD  ($N {\rightarrow} {N\!+\!1}$, $\ket{N+1}$ is a virtual state) and farther into D (${N\!+\!1} {\rightarrow} N^*$), or (ii) an electron tunnels into D ($N {\rightarrow} {N\!-\!1}$,  $\ket{N\!-\!1}$ is a virtual state) and a second electron tunnels from S (${N\!-\!1} {\rightarrow} N^*$). Both paths contribute to the IEC rate. We gather all such processes under the concise notation $N \overset{N\pm 1}{\longrightarrow} N^*$. The IEC energy condition, $|eV_{\rm SD}| \geq \Delta_0$, defines two triangles inside the Coulomb diamond. 
	
	Of particular interest inside the IEC triangles are the sidebands delimited by the excited lines (e.g. regions II+III and II'+III in negative bias). As pointed out above, these regions also allow sequential tunneling through state $\ket{N^*}$. This combination of IEC effective excitation, $N \overset{N\pm 1}{\longrightarrow} N^*$, with sequential transport through the excited state is often labeled as cotunneling-assisted sequential tunneling (COSET) \cite{Wegewijs2001}. In the absence of IEC, other excitation mechanisms are outweighed by relaxation. Therefore,  at a bias below the excitation energy, relaxation annuls the probability of the excited state and hence the sequential tunneling through it; this is why excited lines do not appear at low bias, $|eV_{\rm SD}| < \Delta_0$.

	\section{Interpretation}
	
	\begin{figure}
		\includegraphics[width=\columnwidth]{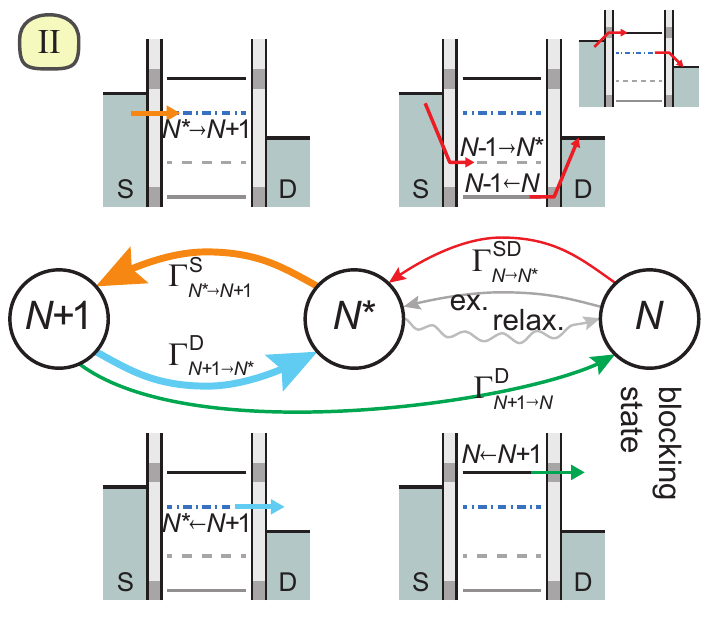}
		\caption{(Color online) Quantum state graph with associated electrochemical-potential diagrams, for device A, region II.  The gray rectangles in tunnel barriers stand for weaker tunneling  via the neighboring transition. $\ket N$ is a blocking state, from which the quantum dot can escape only by low-rate processes (inelastic cotunneling in the second diagram; excitation). Elastic cotunneling is not displayed. In the bottom-right chemical-potential diagram, tunneling into S is equally possible.}
		\label{fig:II}
	\end{figure}
	\begin{figure}
		\includegraphics[width=\columnwidth]{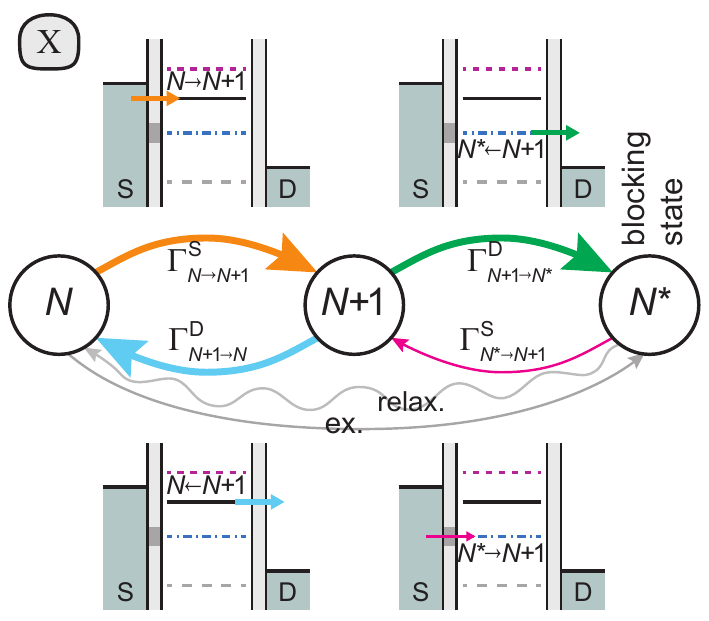}
		\caption{(Color online) Quantum state graph with associated electrochemical-potential diagrams, for device B, region X. $\ket{N^*}$ is a blocking state if the main escape process, namely tunneling from S via transition $N^* \rightarrow N+1$, is hindered (marked with a gray rectangle in the tunnel barrier). Cotunneling transitions are not displayed.}
		\label{fig:X}
	\end{figure}

	We further investigate our noise data within the model, by looking for transitions through which tunneling is relatively weak. These may indicate a delayed-escape path from a reachable state. Such a particular state is referred below as a blocking state.
	
	First, we match the large-Fano-factor data of the two devices onto labeled regions of Fig.~\ref{fig:model}(d). For device A, we expect inside diamond $N$ the existence of regions II, II', III. For device B, we argue that the band below diamond $N$ is region X. The involved processes are summarized in Figs.~\ref{fig:II} and \ref{fig:X}.

	Inside the Coulomb diamond $N$ of device A, IEC triangles are indicated by current and by differential-conductance lines [Fig.~\ref{fig:data}(b),(a)]. The triangles are entirely covered by the COSET sidebands, composed by regions II, II', III. Hence, both IEC and sequential tunneling are present. Fig.~\ref{fig:II} specifically illustrates that IEC breaks the blockade ($N \overset {N\pm 1} {\longrightarrow} N^*$, red arrow), being followed by  sequential tunneling ($N^* \overset{\rm S}\rightarrow N\!+\!1 \overset{\rm D}\rightarrow N^* \overset{\rm S}\rightarrow$ etc., orange and blue arrows). The transport sequence is interrupted by spontaneous relaxation (wavy gray arrow) or by the $N+1\rightarrow N$ transition (green arrow), both resulting in a return to the blocking state, $\ket N$. Because cotunneling is a second-order process, the IEC rate, $\Gamma_{N\rightarrow N^*}^{\rm SD}$, is relatively weak, hence the blocking time is relatively long. Thus, electronic-flow periods alternate with zero-current periods, implying telegraphic transport, i.e. augmented noise. This is indeed what the Fano factor map [Fig.~\ref{fig:data}(e)] reveals. We stress here that the relaxation rate (wavy gray arrow) should be lower than the sequential-tunneling rates (orange and blue arrows) in order allow alternation of current and blocking periods, i.e. super-Poissonian noise.

	Therefore, for device A we have pointed out a blocking state, $\ket N$, connected to other QD states by a low-rate tunneling process, IEC. The IEC rate is usually weak enough to assure an increased Fano factor. In the simple case of all-identical tunneling couplings, and zero relaxation rate, a theoretical value $F=2$ is predicted \cite{Kaasbjerg2015}. Furthermore, $F$ can be substantially boosted by reducing the IEC rate, e.g. when weakening the tunneling via transitions involved in IEC.  The slower IEC escape keeps the QD in the blocking state a longer time and thus increases the transport telegraphicity. Impeded tunneling, resulting in rarer IEC, is graphically suggested in the chemical-potential diagrams of Fig.~\ref{fig:II} by gray rectangles in the two tunnel barriers, at the heights of the $N\leftrightarrow N\pm 1$ transitions.

	Device B exhibits its highest Fano factor below diamond $N$, in a range that we will show corresponds to region X. The bias window of this region contains two transitions, $N^*\rightarrow N+1$ and $N \rightarrow N+1$ (see the chemical-potential diagrams in Fig.~\ref{fig:X}). Like for the other device, we explain, by means of a blocking state, the telegraphic transport signaled by the strong Fano factor. For that, we can suppose that one of the tunneling rates shown in the graph is weak. 
	We assume for the moment that tunneling from source via transition $N^* \rightarrow N+1$ is slow. Provided that the relaxation of $\ket {N^*}$ is slower or of same order, $\ket {N^*}$ becomes a blocking state.
	Leaving the blocking state through the transition $N^* \rightarrow N\!+\!1$ (magenta arrow) or through relaxation (gray arrow) triggers sequential tunneling (the transport sequence $N\!+\!1 \overset{\rm D}\rightarrow N \overset{\rm S}\rightarrow N\!+\!1 \overset{\rm D}\rightarrow$ etc., suggested by orange and blue arrows), ended when the systems falls back into $\ket{N^*}$. As opposed to COSET, the escape path to transport is not a second-order process (IEC), but a weakly lead-coupled tunneling event or internal relaxation. Because the blocking state is due to the weakly coupled tunneling, we are naming this phenomenon sequential tunneling intermitted by weak coupling, SETWEC. As in COSET, the telegraphic character of the transport induces an enhanced Fano factor.

	We now justify for device B the choice of lowering $\Gamma_{N^*\!, N+1}^{\rm S}$, over the other three options in the quantum state graph. For example, if $\Gamma_{N^*\!, N+1}^{\rm D}$ were chosen, then the QD would not have a blocking state\footnote{Nonetheless, were the rate $\Gamma_{N^*\!, N+1}^{\rm D}$ weak, a high-$F$ band would rather appear in the positive-bias domain.} (indeed, it would not stay in $\ket{N+1}$, but would oscillate between $\ket{N+1}$ and $\ket N$). A similar situation would occur if $\Gamma_{N, N+1}^{\rm D}$ were lowered. However, if $\Gamma_{N, N+1}^{\rm S}$ were chosen instead, the QD would have $\ket N$ as a blocking state, instead of $\ket{N^*}$. This statement can immediately be explained by the symmetry of $N$ and $N^*$ in the graph of the QD states [Fig.~3(d)]. The difference between the two candidates is the transition that lifts the blocking state if entering the bias window: $\ket {N^{(*)}}$ is lifted by a transition from this state, namely the source-resonant $N^{(*)}\rightarrow (N+1)^*$. This transition is also the bottom-right edge of the high-$F$ band. Our choice, leading to the edge $N^*\rightarrow (N+1)^*$, is therefore the correct one. We remind that the right corner of region X and that of the measured high-$F$ band are the same bias, $|e|V_{\rm SD} = -\Delta_1$.
	
	Subsequently, we discuss qualitatively the presence of the differential conductance at some boundaries of the studied regions. For device A, we place ourselves in region II. If the bias voltage increases such that transition $N \leftrightarrow N+1$ enters the bias window, one goes outside region II, across the Coulomb diamond edge. In consequence, the quantum state graph (Fig.~\ref{fig:II}) gains one arrow, from $\ket N$ to $\ket{N+1}$. The electronic transport consists now not only of rare cotunneling events (e.g. $N \rightarrow N^*$, red arrow) and sequential tunneling through level $N^* \leftrightarrow N+1$ (loop of orange and cyan arrows), but also sequential tunneling through level $N \leftrightarrow N+1$ (loop formed by the green arrow and the newly added arrow). The orange-cyan loop is faster than the new loop because of higher tunneling rates; indeed, transition $N^* \leftrightarrow N+1$ is more strongly connected to leads than $N \leftrightarrow N+1$. If including the latter in the bias window would only result in the replacement of some faster sequences by slower sequences, then the current would diminish (and the Coulomb diamond edge would exhibit NDC). However, below region II, current grows (meaning positive differential conductance on the Coulomb diamond edge) because the access to the fast, orange-cyan loop solidly increases: in region II, this access is granted by inelastic cotunneling (red arrow) and excitation; below region II, it is substantially raised by transition $N \rightarrow N+1$.
	
	For device B, transport in region X (Fig.~\ref{fig:X}) is carried by fast and slow tunneling sequences (orange-cyan and green-red loops, respectively). If the bias voltage decreases such that the weakly coupled transition, $N^* \rightarrow N+1$, exits the bias window, only the fast, orange-cyan loop remains in the quantum state graph. Hence, each slow tunneling sequence is substituted by several fast sequences. Therefore, the bias voltage decrease results in a current increase, synonym to NDC on the crossed boundary of region X. In a charge stability diagram, this boundary is the line given by the resonance between level $N^* \rightarrow N+1$ and the drain. If the weakly coupled transition were  $N \rightarrow N+1$ instead of $N^* \rightarrow N+1$, then the NDC line candidate would be a different boundary of region X, namely the Coulomb diamond edge; yet, NDC would not occur here, for reasons exposed for device A, region II.

	\section{Numerical simulations}

	\begin{figure*}
	\includegraphics[width=\textwidth]{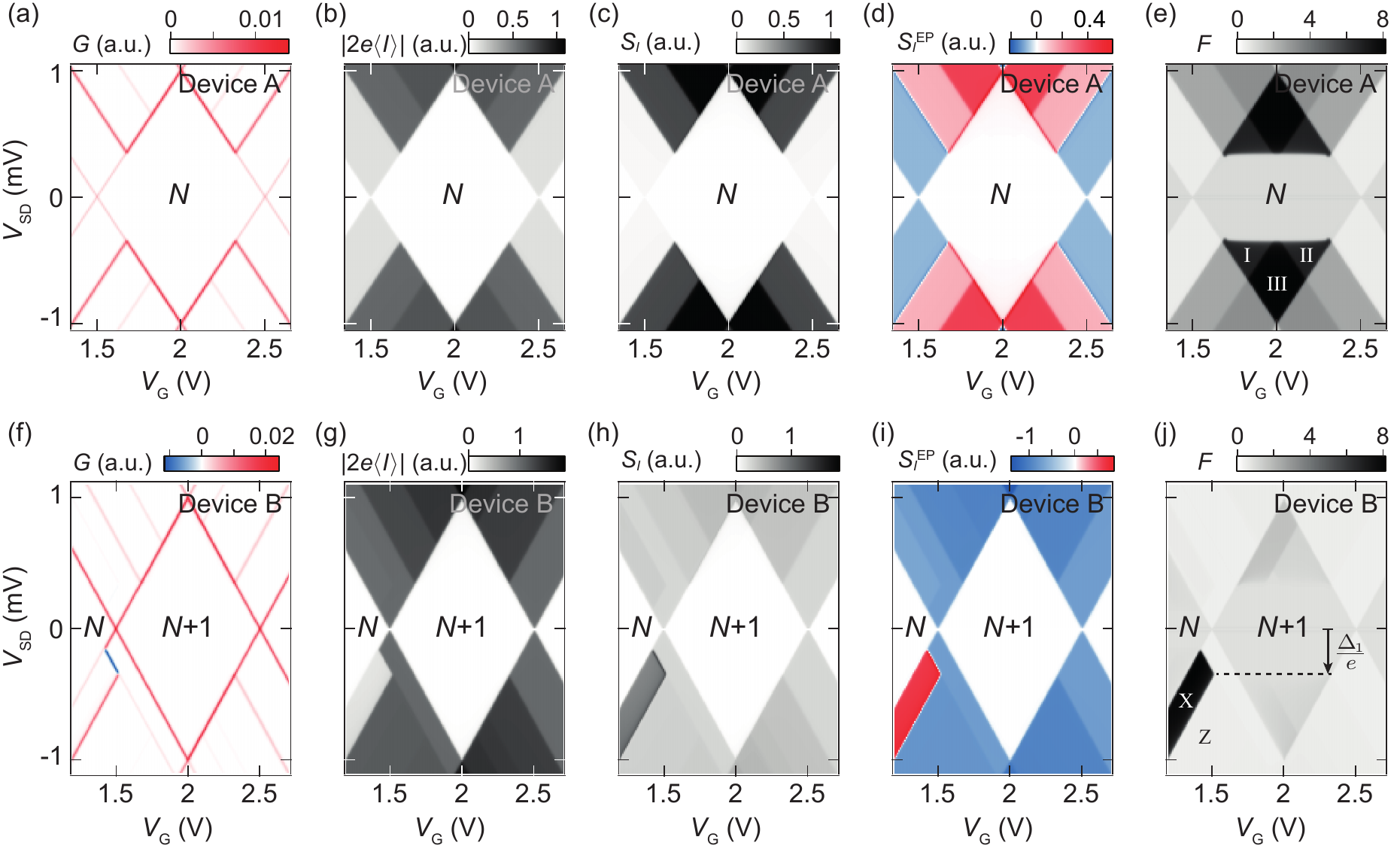}
		\caption{(Color online) Numerical calculations of conductance $G$ and Fano $F$ factor for (a-e) device A ($t_{N, N\pm1}^{\mathrm{S/D}}=0.33$) and (f-j) device B ($t_{N^*, N+1}^{\rm S}=0.1$, $t_{N^*, N+1}^{\rm D}=0.4$, the relaxation rate for the excited state of $N$+1 is $ \Gamma$). All other $t$ amplitudes are 1. The maximal tunneling rates are $\Gamma^{\mathrm{S/D}}=\Gamma=10^{-3}~U_\mathrm{c}/\hbar$ and the relaxation rate is $10^{-3}~\Gamma$. In device A, $F$ is enhanced in regions I, II, III. In device B, $F$ is significantly high in region X, whose top corner is characterized by a bias voltage $-\Delta_0/|e|$ and  right corner by $-\Delta_1/|e|$.}
	\label{fig:simulation}
	\end{figure*}

	In order to validate our interpretations, we run numerical simulations.  To calculate the conductance and the current noise, we employ the master equation approach developed in Ref.~\cite{Kaasbjerg2015}. The master equation describes transitions between the QD states due to (a) first- and second-order tunneling trough the barriers of the QD and (b) internal relaxation and excitation. The transition rates can be found in Ref.~\cite{Kaasbjerg2015}.
	
	For both devices, we simulate three Coulomb diamonds, labeled again as $N-1$, $N$, and $N+1$. The charging energy is $U_{\rm c}  = 1$~meV. We set the temperature $k_B T = 4\cdot 10^{-3} ~U_{\rm c}$.  The maximal tunneling rate is symmetrically chosen, $\Gamma^{\rm S}=\Gamma^{\rm D}=\Gamma_0=10^{-3} ~U_{\rm c}/\hbar$. (A tunneling rate through lead $\alpha$ is proportional to $\Gamma^{\alpha}$; a cotunneling rate is proportional to $\Gamma^{\rm S} \Gamma^{\rm D}$.)
	Tunneling rates are also related to tunneling amplitudes. If $t^\alpha _{N, M}$ is the amplitude of tunneling from lead $\alpha$ into the QD, such that the QD state undergoes the transition $N\rightarrow M$ (see formal definition in Appendix A), then the corresponding tunneling rate is:
	\begin{equation}
	\Gamma_{N\rightarrow M}^{\alpha} \propto \Gamma ^\alpha |t_{N, M}^{\alpha}|^2.
	\end{equation}
	The relaxation rates are $10^{-3}~\Gamma_0$, unless specified. The excitation rates, proportional to the respective relaxation rates, are described by the Bose-Einstein distribution function.
	
	To reproduce the essential features of our data, our calculation utilizes a minimal number of states and a small number of distinct tunneling amplitudes. A more detailed reproduction can be further obtained with more parameters.
	
	For the simulation of device A [Fig.~\ref{fig:simulation}(a-e)], an excited state is introduced at $\Delta_0 = 0.35\cdot U_c$ above the ground state $\ket N$. We weaken tunneling for ground-state transitions ($t_{N, N\pm1}^{\rm S/D}=0.33$, while all other $t$ amplitudes are 1) to obtain high Fano factor values, $F>8$, inside Coulomb diamond $N$ at absolute bias voltages greater than $\Delta_0/|e|$. The maximal value of $F$ is therefore in good agreement with the measurement. The simplicity of the tunneling amplitude set of values wipes out only some nuances of the measured data.

	For the simulation of device B [Fig.~\ref{fig:simulation}(f-j)], an excited state is defined at $\Delta_0 = 0.15\cdot U_c$ above the ground state $\ket N$ and another one at $\Delta_1 = 0.35\cdot U_c$ above the ground state  $\ket {N+1}$. By relatively weakening the source-QD tunneling via transition $N^*\leftrightarrow N+1$ ($t^{\rm S}_{N^*, N+1}=0.1$, versus $t^{\rm D}_{N^*, N+1}=0.4$), an enhanced Fano factor region appears beneath the $N$ diamond and an NDC line sets in at the top of this region, exactly as in the measurement. The emerged high-$F$ band, labeled X, has its rightmost corner placed at $V_{\rm SD}=-\Delta_1/|e|$ and its southeastern edge situated between diamond $N$ and the excited line $N \leftrightarrow (N+1)^*$; the only  transition corresponding to this position is, like in our interpretation, $N^* \leftrightarrow (N+1)^*$. We note that without including the excited state $(N+1)^*$ in the model, this escape process is absent and region X extends infinitely, as shown in Ref. \cite{Kaasbjerg2015}. Further parameters were tuned in the calculation for the secondary purpose of reducing the conductance of the excited line  $N\leftrightarrow(N+1)^*$ resonant with the source: $t^{\rm S}_{N,(N+1)^*}=0.1$.
	
	Obtaining Fano factor values beyond certain thresholds is not possible without an extra ingredient, on which we elaborate here. 
	As derived in Ref.~\cite{Carmi2012}, the Fano factor of the super-Poissonian noise in a telegraphic system is described by the expression 
	\begin{equation}
	\label{eq:FOreg}
	F = 1 +2 \expect n P^2_{\rm off},
	\end{equation}
	where $\expect n$ is the average number of electrons in a sequential-tunneling bunch and $ P_{\rm off}$ the occupation probability of the blocking state. A simple analysis, done in the absence of internal relaxation and excitation, is presented in Appendix B.  It shows that a higher $P_{\rm off}$ is produced when the blocking-state escape rate is lower than the other tunneling rates. $P_{\rm off}$ approaches 1 in the case of a strongly blocking state. The additional ingredient, $\expect n$, is given by the relative strength of the tunneling path that keeps the current on. Concretely, we demonstrate that $\expect n$ is the ratio of the rates illustrated by the cyan and the green arrow respectively (Figs. \ref{fig:II}, \ref{fig:X}). With all tunneling rates equal except for the escape path, Eq.~\ref{eq:FOreg} leads to $F \leq 3$. A simulation of device B confirms the limit $F \simeq 3$ in the special case $P_{\rm off} \simeq 1$ (namely, when tunneling out of the blocking state is extremely low). Our general calculations do reach Fano factors above 3 because of the risen number of charges in a bunch, caused by the tunneling imbalances $t_{N^*, N+1}^{\rm D} / t_{N, N+1}^{\rm S/D} = 1:0.33$ (device A) and $t_{N, N+1}^{\rm D} / t_{N^*, N+1}^{\rm D} = 1:0.4$ (device B).

	\section{Conclusions}
	We identify Fano factor corresponding to markedly super-Poissonian noise in two devices. We propose a model which allows to explain the observed $F>1$ regions in a consistent way. In the exposure of the underlying quantum transport processes, we show that the concept of blocking state is central in the occurrence of enhanced noise. Escaping the blocking state leads to electronic flow 
	for a certain time, until the system falls back into the blocking state. This gives rise to a telegraphic pattern of charge transport, consisting of a random set of charge packages, which determines the enhanced Fano factor. We have identified and proven two mechanisms that can generate telegraphic transport: (i) In COSET (cotunneling-assisted electron tunneling) the blocking state is the ground state of the Coulomb blockade and can be left by cotunneling. If it is fled to an excited state, a transport channel may open. (ii) Outside the Coulomb blockade, when the bias window contains two coupled transitions involving the same charge states, e.g. $N \leftrightarrow N\!+\!1$ and $N^* \leftrightarrow N\!+\!1$. If one state, e.g. $\ket {N^*}$, is weakly coupled, then it becomes a blocking state, causing again telegraphic transport. We term this process sequential tunneling intermitted by weak coupling (SETWEC). SETWEC can be accompanied by negative differential conductance.
	
	\section{Acknowledgements}
	We thank Thomas Hasler for fruitful interactions in the fabrication of device A and Minkyung Jung for measurement guidance.
	
	We acknowledge financial support from the ERC project QUEST and the Swiss National Science Foundation (SNF) through various grants, including NCCR-QSIT.
	
	\section{Appendix A}
	
	In the model, we define the tunnel coupling of the QD to lead $\alpha$ via transition $N \rightarrow M$ the complex amplitude of tunneling between lead $\alpha$ and the QD \cite{Kaasbjerg2015},
	\begin{align}
		\label{eq:M}
		t_{N\rightarrow M}^{\alpha} & = \bra{M} \hat t_\alpha^{\dagger} \ket{N}
		= (t_{M\rightarrow N}^{\alpha})^*,
	\end{align}
	such that  $\ket N$ is the initial state of the QD, $\ket M$ its final state, $\ket N$ and $\ket M$ are consecutive-charge states, and $\hat t_{\alpha}^\dagger$ is the electron creation operator in the QD, coupled to lead $\alpha$. In other words, $\hat t_{\alpha}^\dagger$ is the operator that describes the tunneling of one electron from lead $\alpha$ into the QD. Tunneling rates are proportional to tunneling probabilities, namely the magnitude squared of lead couplings:
	\begin{equation}
		\Gamma_{N\rightarrow M}^{\alpha} \propto |t_{N\rightarrow M}^{\alpha}|^2.
	\end{equation}
	
	We suppose that tunnel couplings do not depend on $V_{\rm G}$ and $V_{\rm SD}$. Moreover, considering only real values, their notation simplifies: $t^\alpha _{N\rightarrow M} = t^\alpha _{M\rightarrow N} =: t^\alpha _{N,M}$.
	
	In a sequential process, two tunneling rates are involved, e.g.
	\begin{equation}
		\Gamma_{N\rightarrow N+1}^{\rm S} \propto |t_{N, N+1}^{\rm S}|^2 {\rm ,~~} 
		\Gamma_{N+1\rightarrow N}^{\rm D} \propto |t_{N, N+1}^{\rm D}|^2.
	\end{equation}
	The tunneling rate of an IEC process, $\Gamma_{N\rightarrow N^*}^{\rm SD} $, accounts for both $\ket{N\!-\!1}$ and $\ket{N\!+\!1}$ as possible virtual states:
	\begin{equation}
		d\Gamma_{N\rightarrow N^*}^{\rm SD} \propto \! \left|\frac{t_{N, N-1}^{\rm D} t_{N-1,N^*}^{\rm S}}{\mu_{N^*\leftrightarrow N-1}-E} + \frac{t_{N, N+1}^{\rm S} t_{N+1,N^*}^{\rm D}}{E-\mu_{N\leftrightarrow N+1}} \right|^2 \!\!\!dE.
	\end{equation}
	
	The $V_{\rm G}$ and $V_{\rm SD}$ dependence of the tunneling rates is explicitly taken into account by means of Fermi-Dirac distribution functions. For details, consult Ref.~\cite{Kaasbjerg2015}.
	
	\section{Appendix B}
	
	The Fano factor in the telegraphic picture can be written as $F = 1 +2 \expect n P^2_{\rm off}$ \cite{Carmi2012}. Here we apply this formula to a specific case and derive expressions for its two ingredients: the probability of the QD to be off, $P_{\rm off}$, and the average number of electrons in a bunch, $\expect n$. A bunch is a package of charge carriers that flow one by one through the QD. The QD is considered to be off when no bunch is flowing through.

	\begin{figure}[!h]
		\includegraphics[width=\columnwidth]{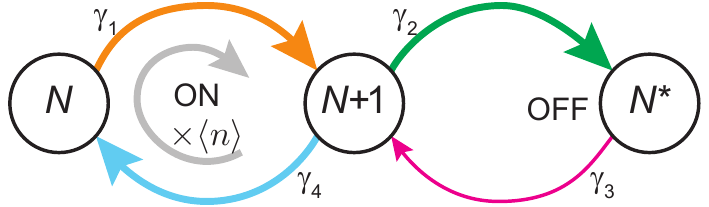}
		\caption{(Color online) Graph of quantum states for a QD that exhibits telegraphic transport. While the QD oscillates in the left loop ($N+1 \leftrightarrow N$), the current is switched on and an average number of $\expect n$ electrons tunnel sequentially from source into the QD and farther into the drain. $\ket{N^*}$ is the blocking state. Cotunneling and internal excitation and relaxation are not taken into account.}
		\label{fig:Apx}
	\end{figure}
	We evaluate a QD with states $\ket N$, $\ket{N^*}$, ${\ket {N+1}}$ and transition rates $\gamma_{1,2,3,4}$, as shown in Fig.~\ref{fig:Apx}. This is a simplified model of device B in region X (compare with Fig.~\ref{fig:X}). We assume that the transport is unidirectional, i.e. the QD is always filled from the source electrode, and emptied into the drain. This is ensured by the bias voltage. The QD exhibits telegraphic transport if, for instance, the escape rate from state $\ket{N^*}$ is much lower than the rates in the left loop ($N+1 \leftrightarrow N$). We formulate a master equation for the probabilities $p_j$ to be in state $\ket j$: 
	\begin{equation}
	\frac d {dt}
		\begin{pmatrix}
		p_N  \\
		p_{N^*}  \\
		p_{N+1}
		\end{pmatrix}
		=
		\begin{pmatrix}
		-\gamma_1 & 0 & \gamma_4  \\
		0 & -\gamma_3 & \gamma_2  \\
		\gamma_1 & \gamma_3 & -\gamma_2 -\gamma_4
		\end{pmatrix}
		\begin{pmatrix}
		p_N  \\
		p_{N^*}  \\
		p_{N+1}
		\end{pmatrix}.
	\end{equation}
 	By solving the master equation in the steady state, $d {\mathbf p}/dt = 0$, we find the occupation probability of state $\ket{N^*}$:
	
	\begin{equation}
		p_{N^*} = \frac {\gamma_1 \gamma_2} {\gamma_1 \gamma_2+ (\gamma_1 + \gamma_4) \gamma_3}.
		\label{eq:Poff}
	\end{equation}
	For an escape rate much smaller than the falling rate, $\gamma_3 \ll \gamma_2$, the off state of the system is equivalent to being in the blocking state: $P_{\rm off} = p_{N^*}$. Eq.~\ref{eq:Poff} already shows that a bigger $P_{\rm off} = p_{N^*}$ is produced by smaller values of the escape rate, $\gamma_3$. With $\gamma_3 \ll \gamma_2$ and $\gamma_4 \lesssim \gamma_1$, one gets 
	\begin{equation}
		P_{\rm off} \simeq 1-\frac{\gamma_4}{\gamma_1} \frac{\gamma_3}{\gamma_2} \simeq 1.
	\label{eq:Poff2}
	\end{equation}
	In this limit, essentially the number of electrons in a bunch, $\expect n$, determines Fano factor.
	
	The average number of electrons in one bunch is
	\begin{equation}
		\expect n = \frac {\gamma_4} {\gamma_2}
		\label{eq:n}
	\end{equation}
	and can be derived as follows: In state $\ket{N+1}$, the probability to fall in the blocking state, $\ket{N^*}$, is $p_b = \gamma_2 / (\gamma_2 + \gamma_4)$, while the probability to go to state $\ket N$ is $\tilde p_b \equiv 1-p_b$. Therefore, the probability that $n$ electrons sequentially tunnel before blocking is $P(n)=\tilde p_b ^n p_b$. The average number of electrons in the sequential-tunneling bunch, $\expect n = \sum_{n=0}^{\infty} n P(n) = p_b \partial_{\tilde p_b} \sum_{n=0}^{\infty} \tilde p_b^n = \tilde p_b/p_b = \gamma_4/\gamma_2$, reads as in Eq.~\ref{eq:n}. In this analysis, the electron that tunnels into the QD and switches it on (${N^*} \rightarrow {N+1}$) is not considered to belong to the consequent bunch.
	In conclusion, under the condition of slow escape from a blocking state, super-Poissonian noise can still substantially grow from the imbalance of tunneling rates related to the pre-blocking state, $\ket {N+1}$.
	
	A limit case arises when tunneling out of $\ket {N+1}$ is balanced: $\gamma_4 = \gamma_2$. In this situation, super-Poissonian noise is characterized by a Fano factor $F = 1+2 P_{\rm off}^2 \leq 3$. If in addition $\gamma_3 / \gamma_2 \rightarrow 0$, then $F \rightarrow 3$.
	
	A similar analysis can be done for a QD described by the state graph of Fig.~\ref{fig:II}, in the absence of cotunneling and internal excitation and relaxation.

\end{document}